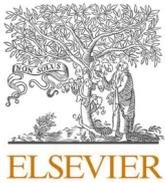

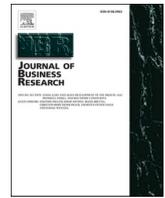

# How public funding affects complexity in R&D projects. An analysis of team project perceptions

José M. González-Varona [*], Natalia Martín-Cruz, Fernando Acebes, Javier Pajares

*GIR INSISOC - Department of Business Organization and Marketing and Market Research, University of Valladolid, Valladolid 47011, Castile and Leon, Spain*

| ARTICLE INFO | ABSTRACT |
|---|---|
| *Keywords:*<br>R&D project<br>Public-funded project<br>Complexity<br>Project management<br>SME<br>Case study | In this paper, we apply a case study approach to advance current understanding of what effects public co-funding of R&D projects have on project team members' perceived complexity. We chose an R&D project carried out by an industrial SME in northern Spain. The chosen research strategy was a qualitative approach, and sixteen employees participated in the project. We held in-depth semi-structured interviews at the beginning and end of the co-funded part of the project. NVivo data analysis software was used for qualitative data analysis. Results showed a substantial increase in perceived complexity. We observed that this was due to unresolved tension between the requirements of the project's co-financing entity and normal SME working procedures. New working procedures needed to be developed in order to comply with the co-financing entity's requirements. However, overall perceived complexity significantly decreased once the co-financed part of the project was completed. |

## 1. Introduction

Today we are witnessing ever-increasing projectification of the economy. A paradigm shift is taking place in companies in which projects are no longer adjacent to operations, but are becoming the fundamental tool for doing work and solving problems (Geraldi & Söderlund, 2018; Project Management Institute, 2020; Schoper, Wald, Ingason, & Fridgeirsson, 2018). According to World Bank data (Bank, 2019), the gross domestic product (GDP) generated annually through project delivery accounts for approximately 24% of total global GDP ($23 trillion), measured in gross capital formation terms (Scranton, 2015). However, these estimates are rough, because a valid universally accepted measure of the degree of projectification still does not exist (Schoper et al., 2018). All sectors of the economy are increasing project-based work, which will have a significant impact on creating millions of jobs in the years ahead and will create an ever-greater gap between the need for skilled project management workers and their availability (Project Management Institute, 2017). In addition, projects are essential for research, innovation, and organisational change.

Consequently, project management as a discipline has grown significantly in recent decades, although this is not necessarily related to increased project performance. Many projects fail to meet some of their strategic objectives, goals, or participants' expectations (Flyvbjerg, 2016; Meredith & Zwikael, 2020; AJ Shenhar & Dvir, 2007), with perceived complexity being one of the elements that impacts a project's success (Bakhshi, Ireland, & Gorod, 2016; Maylor & Turner, 2017; Maylor, Vidgen, & Carver, 2008; Tatikonda & Rosenthal, 2000). Both perceived complexity and project success are multidimensional concepts, and increasing complexity can negatively influence project performance. However, how this influences the project's success has yet to be determined, and academic studies are needed in order to gain an in-depth understanding of this impact (Bjorvatn & Wald, 2018; Bosch-Rekveldt, Bakker, & Hertogh, 2018).

Small and medium-sized enterprises (SMEs) increasingly use participation in publicly funded collaborative R&D projects to develop new products, technologies, or business lines. Despite the existing academic literature which recognises that participation in these projects generates positive impacts in innovation terms, current knowledge concerning the complexities generated, and the managerial responses that need to be given to these complexities remains limited (González-Varona, López-Paredes, Poza, & Acebes, 2021; Pajares, Poza, Villafañez, & López-Paredes, 2017). Maylor et al. (2013) synthesised project complexity on three dimensions which project managers must address from a subjective approach: structural complexity, socio-political complexity, and emergent complexity.

When project complexity increases due to public funding, the balance between reducing the risk in an R&D project (by raising public funds) and the increased complexity (given the bureaucracy associated

* Corresponding author at: University of Valladolid, Pº. Prado de la Magdalena, s/n, Valladolid, Castile and Leon 47011, Spain.<br>*E-mail address:* josemanuel.gonzalez.varona@uva.es (J.M. González-Varona).

with public agencies) becomes a challenge for companies. We therefore formulate the following research questions:

RQ1 - How does the project team perceive complexity in publicly co-funded projects in SMEs?

RQ2 - What elements of complexity are perceived more intensely at the beginning of the project and after completing the part of the project co-financed by a public entity?

RQ3 - Does complexity vary homogeneously before and after completing the part of the project co-financed by a public entity?

In order to bridge the gap, we consider a case study approach to be appropriate for our research. The case chosen is an R&D project to be carried out by an industrial SME in the Castilla y León region of Spain, which was co-financed by a public entity in its first phase. A qualitative approach was chosen as a research strategy to obtain the necessary data. To collect data, semi-structured interviews were held with the project participants. For the subsequent qualitative analysis of the data obtained during interviews, NVivo data analysis software was used (version 1.6).

This paper aims to contribute to the academic literature on how the co-funding of an R&D project by a public entity influences internal project participants' perceived complexity. More specifically, we explore how project participants interact and work with one another, and we determine how the organisation's informal structure changes –which has significant effects on how work is done and how procedures and processes change. Changing the way project participants interact leads to the emergence of new structures and features that may affect SME culture. In addition, this research will help to improve business practice via a better understanding of the complexity of projects co-financed by public entities so as to thus enhance project performance. Practitioners will be able to obtain useful information to increase the success of their co-financed projects.

In our research, we studied how complexity was perceived by internal participants in an R&D project undertaken in an SME, and co-funded by a public entity in two of its stages. The study analysed in depth: (1) perceived complexity at the beginning of the project; (2) perceived complexity at the end of the part of the project that was co-funded, but not fully completed.

The rest of the paper is structured as follows. Section 2 discusses the theoretical background as well as studies related to the evaluation of complexity performance in the context of projects. Section 3 describes the research work. Section 4 deals with analysis of the data. Section 5 presents the results of this study. Section 6 discusses the results obtained. Finally, the main conclusions to emerge from this research as well as directions for future enquiry are presented.

## 2. Theoretical framework

### 2.1. Complexity in the project management context

According to Grabher (2002), projects can prove complex to manage. Indeed, complexity in the project context is one of the most critical factors that can impact a project's success. Project complexity has grown steadily over the last few decades (Geraldi et al., 2011; T. Williams, 1999). Williams (1999) pointed to the ever-increasing complexity of products being developed and the ever-shortening timeframes as being two of the fundamental causes that increase project complexity. Increasing complexity has fostered the interest of academic research in this field (Marnewick et al., 2017; Tatikonda & Rosenthal, 2000; Vidal & Marle, 2008).

The term "complexity" used by academics is a narrower concept than is employed by practitioners; in fact certain context-related aspects that practitioners point to as being complex are identified by academics as complicated (Baccarini, 1996; Girmscheid & Brockmann, 2008; Remington & Pollack, 2016). This is because theoretical complexity focuses on emergence, uncertainty, non-linearity and interdependence or connectivity among the elements present in a project.

As indicated by Vidal & Marle (2008) and Vasconcelos & Ramirez (2011), the term complexity in projects has led to confusion, with various arguments about what is "complex" or "complicated". For the purposes of this case study, we do not distinguish between the terms "complex" and "complicated" –following the common usage employed by several authors (Geraldi, Maylor, & Williams, 2011; Vidal & Marle, 2008). In both rounds of interviews, practitioners' responses can be seen to include both the terms "complexity" and "complicated". According to Maylor and Turner (2017, p. 1,077), "one person's complex is another's complicated".

Complexity will impact project goals and objectives, project planning and organisation as well as staff recruitment requirements. San Cristóbal et al. (2018) indicate that complexity in the project context has become the focus of attention for several reasons: (a) it impacts the way the project is planned, executed and controlled; (b) it can hinder the identification of goals and objectives; (c) it also influences how the project is organised as well as the skills required by workers; (d) it can impact project objectives (scope, time, cost, risks, etc.).

Complexity can positively influence projects by increasing the probability of success or, on the contrary, can negatively influence project success (Project Management Institute, 2013; San Cristóbal et al., 2018). Complexity can lead to the emergence in projects of new elements that are not predictable merely through a knowledge of the behaviour and interactions between project elements. This can generate opportunities and, therefore, have a positive effect. It can also lead to negative effects, which arise from new threats and that increase the difficulty involved in understanding and controlling projects. Project managers need to manage complexity by taking advantage of and reinforcing opportunities whilst also reducing any threats which may trigger the adverse effects of complexity (Bubshait & Selen, 1992; San Cristóbal et al., 2018; Vidal & Marle, 2008). The similarity between risk management and complexity is evident. However, the complexity approach includes a component of subjectivity that allows it to incorporate more options than those identified when risks are recorded (Maylor & Turner, 2017).

Historically, scientific research has adopted a dual approach to complexity in project management. According to (Cicmil, Cooke-Davies, Crawford, & Richardson, 2009; Schlindwein & Ison, 2004; Vidal & Marle, 2008) we can distinguish between two approaches to complexity: descriptive complexity and perceived complexity. A rationalist approach takes complexity to be an intrinsic property of the project, which is mainly technological and organisational and which can therefore be measured and quantified. On the other hand, a subjective approach, which considers complexity to be a perception of the individuals involved in projects, will depend on the experiences lived during the project's life cycle. This is also known as "complexity of projects".

The approach to complexity as a "lived experience" by project participants has been used by several authors as the most suitable way to study complexity in the project management field (Baccarini, 1996; Geraldi et al., 2011; Maylor & Turner, 2017; Maylor et al., 2013; Vidal & Marle, 2008). Even though both approaches are applicable to project complexity, in practice the problem lies in the fact that project managers cannot encompass all the complexity generated in projects (Vidal & Marle, 2008). Our paper therefore focuses on complexity as perceived by project participants, i.e., a subjective approach to complexity.

According to Baccarini (1996), one definition of project complexity is that it consists "of many varied interrelated parts". He advocated implementing it in terms of the differentiation and interdependency of varied elements. Terry Williams (1999) published a paper in which he attempted to answer the question "what constitutes project complexity?" by taking the paper published by Baccarini as a starting point. In the paper, he identified two dimensions of project complexity: structural complexity and uncertainty. In addition, structural complexity has two sub-dimensions: the number and interdependence of project elements, such as tasks, specialists, components. He also proposed two sub-dimensions of the uncertainty dimension: uncertainty in goals and means.

Geraldi et al. (2011) systematically reviewed the relevant academic literature published to date, and constructed an integrated framework to assess the complexity of projects. Five dimensions of "complexity of projects" were identified: structural, uncertainty, dynamics, pace, and socio-political. Projects exhibit a mixture of these dimensions and these are often interdependent on one another. With this paper, the authors demonstrate complexity as a "lived experience" by project managers. Therefore, the types of complexity that project managers will identify in a project will have a subjective component.

The five-dimensional model of Geraldi et al. (2011) was followed by Maylor et al. (2013) who constructed the Complexity Assessment Tool (CAT) model, which summed up in three dimensions the complexity of projects from a subjective approach: structural complexity (including pace); socio-political complexity and emergent complexity (including uncertainty and dynamics).

Structural complexity is the easiest for practitioners and researchers to identify, and it increases with size, variety, breadth of scope, level of interdependence between people or tasks, pace or variety of work to be done, number of locations, and time zones. The existence of strict deadlines is a source of complexity because it leads to an increase in work pace and people's stress (Geraldi et al., 2011; Maylor & Turner, 2017; Terry Williams, 2017).

Socio-political complexity is related to the project's importance to the organisation, people, power and politics of all the parties involved in both the project team and externally. It increases when there is lack of agreement or compromise between the parties involved and the level of politics or power play that the project is subject to. These are project objectives that are not shared or which do not fit strategic goals, personal objectives or priorities that come into conflict with those of the project (Geraldi et al., 2011; Maylor & Turner, 2017; Swart, Turner, Maylor, Prieto, & Martín-Cruz, 2017).

Finally, emergent complexity depends on two related factors: uncertainty and change. A new technology or process will increase uncertainty, as will a team's lack of experience or insufficient information (Maylor, 2013). Additionally, changes in the requirements involved, the technology or parties will increase emergent complexity. Emergent complexity increases when objectives, vision or success criteria are not well defined, and uncertainty arises in their interpretation. Emergent complexity is related to risk. Uncertainty is one of the characteristic elements of risk, i.e., of an uncertain event which, should it occur, will affect project objectives (Geraldi et al., 2011; AJ Shenhar & Dvir, 2007; Terry Williams, 2005).

We use these definitions of complexity to describe the nature of perceived complexity in an SME's R&D project, which we selected as a case study. We explain why we selected a case study in Section 3: Methodology.

### 2.2. Complexity for publicly-funded R&D projects

SMEs need to develop and exploit new knowledge in order to maintain their technological competitiveness. One increasingly used strategy is to collaborate in strategic technology alliances with other organisations, with participation in publicly funded R&D projects being one widely employed option. According to Radas et al. (2015) and Spanos et al. (2015), public funding of R&D projects aims to encourage research in high-risk and complex projects, which would otherwise be difficult for companies to carry out without institutional support. It has a multiplier effect on R&D spending in the private sector, and fosters the most basic and radical innovation responsible for the new products that come on to the market (Clausen, 2009; Hottenrott & Lopes-Bento, 2014; Martín-Barrera, Zamora-Ramírez, & González-Ramírez, 2017). This offers an obvious advantage for a firm because co-financing with public funds limits the economic risk the firm must face given that a project of this size requires significant investment. According to Meuleman and De Maeseneire (2012), it will make obtaining private financing from banks and venture capital firms easier in the future.

However, R&D funded projects entail additional complexity as a result of having to perform specific bureaucratic tasks as well as the intrinsic uncertainty that characterises them. In addition, complexity has increased due to certain factors, such as cross-functional teams, globalisation and shorter product life cycles (Chronéer & Bergquist, 2012). It should be noted that, although these projects significantly increase structural and socio-political complexity, emergent complexity is the most important because change is an intrinsic element of R&D projects (Maylor et al., 2013). As indicated by Remington and Pollack (2016), the uncertainty caused by a lack of knowledge of design requirements or technical aspects means that technical complexity is also a characteristic of R&D projects. Furthermore, publicly funded R&D projects add new requirements that increase complexity: for example, in EU-funded projects, the monitoring mechanisms in place, which require compliance with strict project control procedures, such as regular reporting, red tape, etc., and which reduce flexibility and create additional rigidities that particularly impact SMEs (Matt, Robin, & Wolff, 2012).

## 3. Methodology

We chose a qualitative research case study as a methodology. According to Yin (2009), case studies are suitable for researching "a contemporary phenomenon in depth and within its real-life context, especially when the boundaries between phenomenon and context are not clearly evident". A case study allows in-depth research and the identification of processes that may go unnoticed (Eisenhardt, 1989). In addition, we noted that case studies are required in the scientific literature in order to investigate institutional factors that may influence projects (Orr & Scott, 2008). Our case study thus contributes to theory development by supporting the governance responses that we identify.

In addition, a qualitative approach was chosen as a research strategy in order to obtain the necessary data for our study. This strategy has been used when studying a specific group or population. According to Creswell and Poth (2016), the qualitative approach is useful for obtaining detailed and complete knowledge of the study population through interviews or surveys, which allow us to collect data by directly interacting with the study population.

To perform this analysis, we considered the conditions of the company, stakeholders, and the context, among other aspects. In order to capture the complexity of the problem under study, semi-structured interviews were held to obtain relevant data from well-informed interviewees. According to Yin (2009), in case studies, conducting interviews is a key source of evidence because most interviews deal with human issues or actions.

Case studies have frequently been used by researchers to study complexity of projects in different activity sectors and to guide practitioners' practice in future projects (Qiu, Chen, Sheng, & Cheng, 2019; A. J. Shenhar, Holzmann, Melamed, & Zhao, 2016; Aaron Shenhar & Holzmann, 2017; Turner, Aitken, & Bozarth, 2018). Some examples of case studies into the complexity of technological innovation projects include (Bosch-Rekveldt et al., 2018; Potts, Johnson, & Bullock, 2020; Poveda-Bautista, Diego-Mas, & Leon-Medina, 2018).

In our paper, we analysed the "Smart"[1] project of the industrial SME "Pressure" as a case study because it was a publicly-funded R&D project of an SME in which we had access to the project team and the company's management as part of a research project involving a variety of both public and private participants. This availability allowed us to understand where complexity arises and how project team members respond to these situations.

---

[1] The research case has been anonymised. The names of the project and the company have been changed to ensure confidentiality.

### 3.1. The "smart" project

The aim of the "Smart" project was to design and build a prototype of a Pressure-Gas-Temperature test bench using Hot Isostatic Pressing (HIP) technology, with far superior performance to that available on the market. The project was led by the company "Pressure", which put its previous knowledge of high pressures to best use in so-called "High-Pressure Processing" (HPP) technology to investigate the compaction and cleaning of materials at high temperature (up to 2,000 °C) in an inert atmosphere at high pressure (up to 2,000 bar) with HIP technology. Specifically, it was intended to be applied to materials obtained through additive manufacturing.

"Pressure" is an SME in the technology sector in Castilla and León (Spain) with subsidiaries in the United States and Mexico. It stands out for its innovative character and R&D projects. In 2018, it started an innovative machinery business line based on new technology for the company: HIP.

The project started in October 2016 and ended in December 2020. The development part of the prototype was completed in October 2019, when it was presented to the public during a public event held in November 2019. The marketing phase was completed from that date to December 2020. Setting deadlines for completion made it possible to evaluate the results at specific time points. Through the "Smart" project, "Pressure" started a new strategic business unit that focused on manufacturing machines for post-processing metal parts, and which still continues today as the "Pressure" website shows, which offers its products and services with HIP technology.

The "Smart" project budget (approx. €2 m) was supported by public funding from the European Regional Development Fund (ERDF), managed by a regional business competitiveness body. Co-financing limited the economic risk taken by "Pressure" but increased the administrative and bureaucratic burden by forcing it to revise its internal working procedures, which reduced process flexibility. Public funding was granted for the development part of the machine prototype, which had an established execution deadline that started in May 2019 and ended in October 2019.

To undertake the project, "Pressure" collaborated with several external organisations: two public universities, two technology centres in north Spain, a regional public body and a company in the Basque Country (northern Spain). Internally, the project was led by a team of 16 employees from different company departments. Unlike other previous projects, "Pressure" had to face not only the novelty involved in HIP technology, which had not been used before, but also the search for new customers, the development of a new product and the co-financing of the project with public funds, which generated additional complexity.

### 3.2. Data collection

Data collection began in 2019 when we gained access to the employees who formed part of the "Smart" project team. The project team included the "Pressure" manager, the project manager and five department managers: administration, R&D administration, R&D, operations and production engineering, plus other participants from the R&D, operations and production engineering departments. In all, there were 16 employees involved, summed up in Fig. 1.

Data collection was conducted through semi-structured interviews held with project participants and which lasted approximately 1.5 to 2 h with each participant. Data collection objectives were to identify not only the complexities perceived by team members before and after the end of the project's public funding, but also the responses provided.

#### 3.2.1. Semi-structured interviews

Between 2019 and 2020, we made good use of the 'Smart' project team's availability to obtain information by holding semi-structured interviews. First, between March and May 2019, we interviewed the 16 project team members to record the complexities they identified at the start of the project, and to learn their daily work processes and communication procedures in the 'Smart' project.

Second, in October 2019 when the prototyping of the new machine and public funding had been completed and before starting the project's commercial phase, project participants were again interviewed in order to identify any new perceived complexities. They were also asked about the complexities they had perceived while the prototype was being constructed.

The interview protocol was based on the Complexity Assessment Tool (CAT) developed by Maylor et al. (2013), and by focusing on complexity and organisational practices in publicly funded R&D projects. It was adapted to the specific project characteristics selected for the case study. CAT comprises a generic questionnaire applicable to a wide range of projects, including technological innovation projects. This tool allows managers to assess the complexity of projects on three identified dimensions: structural, socio-political, and emergent (Maylor

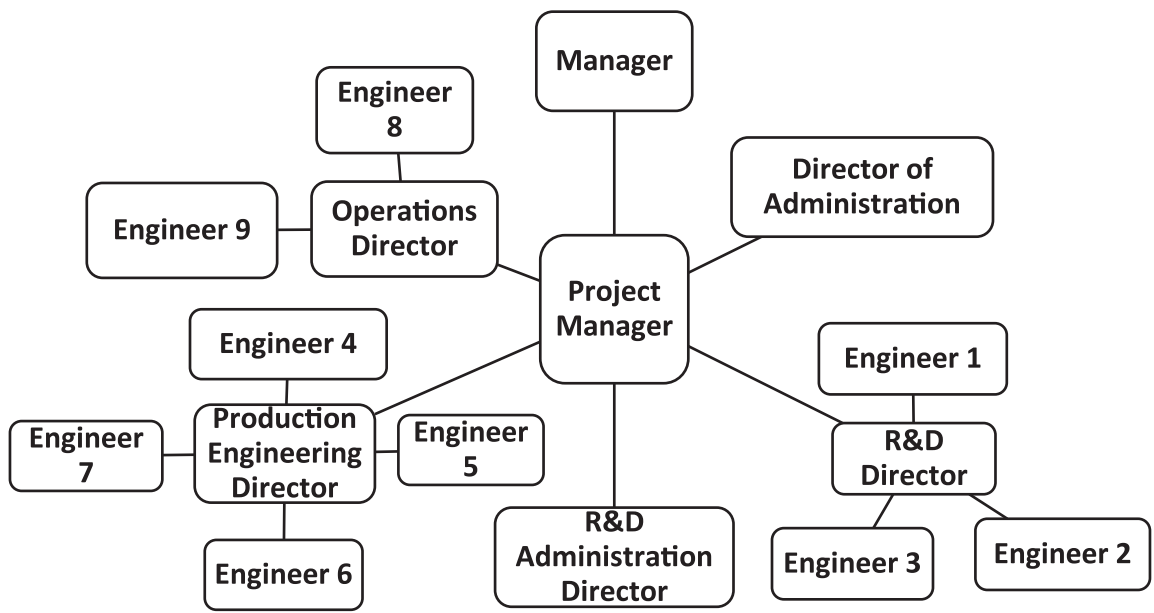

**Fig. 1.** Participants in the "Smart" project. Source: the authors.

et al., 2013).

Interviews were held with project team members in a semi-structured manner following the established protocol, and they provided insight into where complexity arose and what relations existed. The interview protocol contained a sequence of questions in which interviewees were initially required to give an overview of the project and to identify perceived complexities. They also had to indicate the responses they planned to give to the perceived complexities. Finally, interviewees were asked about their opinion on customer relationship and commitment to the project.

The use of semi-structured interviews allowed us to further develop the answers given by the interviewees, who could explain or expand on information whenever they considered it necessary. This allowed us to confer greater consistency on the data collected because semi-structured interviews provide for a better interpretation of the answers given that interviewees can use different language or ideas to refer to common concepts. This also allowed us to focus the interviews on new aspects that were not considered *a priori*, but which proved relevant to the research carried out.

Text quotations from the interviewees are included in the Results section in order to convey the richness of the interview data in greater detail (Saunders, Lewis, & Thornhill, 2009). All the interviews were held with a minimum of two interviewers and were fully audio-recorded for transcription purposes. The semi-structured interview approach facilitated the interviewers' relationship with the interviewee and the conversational nature of the interviews. The data obtained were used for the qualitative analysis. In addition –and in order to keep the research conducted during the project secret– the principal investigator reached an oral confidentiality agreement with the company manager and team members. The "Pressure" manager did not consider it necessary to sign a paper agreement because of the good relationship with the researchers. In addition, all the information on the "Smart" project development that was generated while the project was underway was shared by the whole project team, as was information about the research process on perceived complexity. As a result, we obtained a detailed and relevant dataset for our research.

## 4. Data analysis

In order to investigate the nature of internal participants' perceived complexity, we chose qualitative methods to collect valuable data from respondents. We interviewed all the participants who held responsibilities and who were involved in the project at two time points: before and after the R&D-funded project's formal deadline date. Pseudonyms were used to preserve anonymity. Each interview lasted approximately 1.5 h and was fully transcribed for NVivo analysis. Version 1.6 of the NVivo software was used for qualitative data analysis. The data obtained from all the interviews were coded according to the complexity dimensions identified by (Maylor & Turner, 2017; Maylor et al., 2013; Turner et al., 2018): structural, socio-political, and emergent.

For better data management purposes, we created sub-codes that grouped those words which represented the complexities identified. For the structural complexity dimension, we thus created the sub-codes: home, department, deadline, budget, resources, and partners. This complexity increases with the number of people involved, the budget used, the number of internal and external interrelationships, the variety of the work to be done, the pace and scope of the work to be done, and the number of departments involved. On the socio-political dimension (communication, trust, coordination, and tension), it increases with the difference in the people involved, hierarchy levels, lack of commitment, lack of knowledge about project objectives and conflicts between stakeholders. The emergent dimension (customer, crisis, differences, uncertainty, facilities, new, problems, delays, and technology) increases with the novelty of the project and when technological and commercial maturity are lacking, which may result in problems being perceived in the project. The same will also occur when there is a lack of previous experience, a lack of information or when any changes are imposed (Maylor & Turner, 2017).

As a result, we obtained a dataset made up of the data collected during interviews, which represented the complexities identified by project team members (see Table 1). These data allowed us to identify the main complexities pinpointed and to compare the two most relevant time points: the start and the formal completion of the publicly funded R&D project.

## 5. Results

We analysed the complexity perceived by senior management and team members at the beginning and the end of the publicly funded project. With the results obtained from the analysis, we built Table 1, which reflects the following data: the first two columns include the number of times a specific complexity was mentioned by those involved in the project, both before and after its completion. In the next two columns, we note the percentages of perceived complexities, both before and after project completion. This provides us with a measure of the magnitude of each perceived complexity, both before and after completing the co-funded portion of the project. The last column contains the percentages of complexity from the first and second rounds of interviews in order to obtain total perceived complexity.

If we compare the complexity perceived by project participants during the first round of interviews to that perceived during the second round, we observe that the complexity perceived in the first round is much greater than is perceived during the second round. In fact, of the 100% perceived complexity during the two rounds of interviews, 72% was perceived during the first round and 28% during the second.

As regards perceived complexity during the first round of interviews, we found that most of the interviewees in the "Smart" project perceived emergent complexity as the most relevant. Structural and socio-political complexities appeared in smaller numbers. During the second round of interviews, the order of relevance was repeated in terms of the type of perceived complexity, but to a significantly smaller degree.

Perceived complexities depend on the different departments and types of involvement in the project (Maylor et al., 2013; Schlindwein & Ison, 2004). The "Pressure" manager detected a higher degree of structural complexity when the project began, "… although, technically speaking, the project was a complex challenge, organisationally it was a challenge, perhaps for us one that is more difficult to organise …".

### 5.1. First round of interviews (before the project grant period)

During the first round of interviews (Table 2), emergent complexity was detected to a greater degree, followed by structural complexity, and then socio-political complexity. In emergent complexity, the complexity related to new customers stood out, followed by the problems identified, and then the novelties pinpointed. In structural complexity, department complexity was the most important. The complexity related to a strict deadline for the end of the subsidised period was also especially important. In socio-political complexity, the need for communication was perceived as the most widely identified complexity by the project team –specifically the three departments in charge of performing project tasks. The need for agreements was also a source of complexity found during the first round of interviews.

When we looked at the complexities perceived by the different departments, we noticed significant differences in both complexity type and the number of references found in each department. In all the departments, emergent complexity was the most commonly reported complexity, followed by structural complexity, and socio-political complexity. The greatest source of complexity in these departments was related to the perceived problems arising from engaging in a project that was totally different to the ones undertaken to date.

The operations, R&D and product engineering departments

**Table 1**
Comparison of perceived complexity before and after completing the publicly funded project.

| Perceived complexity | Interviews before completion | Interviews after completion | Interviews before completion | Interviews after completion | Total percentage |
|---|---|---|---|---|---|
| Emergent | 401 | 142 | 74% | 26% | 100% |
| Customer | 109 | 29 | 79% | 21% | 100% |
| Crisis | 3 | 2 | 60% | 40% | 100% |
| Differences | 7 | 5 | 58% | 42% | 100% |
| Uncertainty | 0 | 7 | 0% | 100% | 100% |
| Facilities | 4 | 6 | 40% | 60% | 100% |
| New | 63 | 13 | 83% | 17% | 100% |
| Problems | 170 | 63 | 73% | 27% | 100% |
| Delays | 11 | 12 | 48% | 52% | 100% |
| Technology | 34 | 5 | 87% | 13% | 100% |
| Structural | 165 | 85 | 66% | 34% | 100% |
| Home | 12 | 10 | 55% | 45% | 100% |
| Department | 88 | 29 | 75% | 25% | 100% |
| Deadline | 30 | 18 | 63% | 38% | 100% |
| Budget | 8 | 6 | 57% | 43% | 100% |
| Resources | 8 | 21 | 28% | 72% | 100% |
| Partners | 19 | 1 | 95% | 5% | 100% |
| Socio-political | 77 | 23 | 77% | 23% | 100% |
| Communication | 37 | 12 | 76% | 24% | 100% |
| Confidence | 27 | 8 | 77% | 23% | 100% |
| Coordination | 7 | 2 | 78% | 22% | 100% |
| Tension (Stress) | 6 | 1 | 86% | 14% | 100% |
| Total complexity | 643 | 250 | 72% | 28% | 100% |

*Source:* the authors.

**Table 2**
Complexity perceived per department during the second round of interviews.

| Perceived complexity | Operations | R&D | Production engineering | Departments Administration | Project manager | Manager | R&D administration |
|---|---|---|---|---|---|---|---|
| Emergent | 103 | 85 | 112 | 13 | 43 | 20 | 25 |
| Customer | 30 | 27 | 24 | 7 | 13 | 6 | 2 |
| Crisis | 3 | 0 | 0 | 0 | 0 | 0 | 0 |
| Differences | 1 | 3 | 1 | 0 | 2 | 0 | 0 |
| Uncertainty | 0 | 0 | 0 | 0 | 0 | 0 | 0 |
| Facilities | 2 | 2 | 0 | 0 | 0 | 0 | 0 |
| New | 21 | 6 | 16 | 0 | 6 | 0 | 14 |
| Problems | 42 | 43 | 60 | 6 | 3 | 14 | 2 |
| Delays | 2 | 0 | 5 | 0 | 0 | 0 | 4 |
| Technology | 2 | 4 | 6 | 0 | 19 | 0 | 3 |
| Structural | 58 | 22 | 44 | 13 | 17 | 7 | 4 |
| Home | 6 | 2 | 3 | 1 | 0 | 0 | 0 |
| Department | 33 | 10 | 21 | 8 | 14 | 0 | 2 |
| Deadline | 10 | 5 | 11 | 1 | 1 | 1 | 1 |
| Budget (money) | 3 | 0 | 5 | 0 | 0 | 0 | 0 |
| Resources | 4 | 0 | 3 | 0 | 1 | 0 | 0 |
| Partners | 2 | 5 | 1 | 3 | 1 | 6 | 1 |
| Socio-political | 23 | 12 | 12 | 9 | 8 | 7 | 6 |
| Communication | 14 | 8 | 2 | 7 | 0 | 2 | 4 |
| Trust | 6 | 3 | 8 | 0 | 5 | 3 | 2 |
| Coordination | 0 | 1 | 0 | 2 | 3 | 1 | 0 |
| Tension (Stress) | 3 | 0 | 2 | 0 | 0 | 1 | 0 |

*Source:* the authors.

identified the greatest number of complexities of the three types in the project. The operations department was the one which pointed to the greatest amount of total complexity.

It should be noted that the administrative part of the project was divided into two departments: one corresponding to the firm's general administration, and the other to R&D administration. The latter is a specific department created by direct order of the company manager to relieve the other departments involved in the project's administrative tasks, and to ensure justification of the documents required by the ERDF. Particularly worth highlighting is the emergent complexity reported by the R&D administration department.

In May 2019, when the co-funded part of the project commenced, the complexity perceived by the manager became more socio-political –mainly due to leadership and "ego" conflicts. As the manager points out, "… this is a very specific project that needs a different way of doing things and a complete change of mindset…".

Unlike the "Pressure" manager, the project manager perceived emergent complexity to be the most relevant, due to the new technology to be used. He also perceived the structural complexity deriving from the administrative burden that managing a project co-financed with public funds would entail. It is striking that he did not detect any complexity of a socio-political nature.

The perceived complexity originating from public funding through the ERDF was mostly structural and related mainly to the project's public co-financing as well as a deadline being set by which the project had to be completed and justified. Given these two conditions, the usual operational procedures had to be adapted to meet the funding body's requirements because it was necessary to provide documentary evidence that the project would meet the requirements of the public body co-financing the project. Four employees indicated that time had to be spent on non-project work. As indicated by the project manager, "… I spent one morning here for something that would not add much value to the project –paper work – but you know that you have to do it that way …". As a source of structural complexity, six employees identified the accumulated delay in the project execution deadlines because, if they started their activity after the established deadline, they would fail to meet it and delays would accumulate. As one of the engineers from the production department pointed out "… there is also the issue of deadlines, i.e., nobody wants to be the last one, in inverted commas, on whom the project depends, right?".

It was not only structural complexity that had an influence, because the perceived socio-political complexity arising from the interdepartmental coordination and collaboration needed to meet the ERDF requirements for project justification was also perceived. As the R&D administration manager put it, "… so that's a difficult thing; getting everyone involved in something that is the company's business and not just specifically one individual's…". Many of the people involved in the project, who focused on their day-to-day activities, attached secondary importance to the tasks needed to meet the requirements for justifying the project to the ERDF. In addition, new procedures had to be established and standardised in order to meet the justification requirements, although many people did not realise the need for more coordination between departments –which led to tension. According to the administration director, "… we are trying to establish different procedures and are trying to automate them because there are many of us involved, given that an important part of the project is the administrative management of everything related to the project. Yet in the technical part, they think that we do not care about the technical part of the project …". Tensions arose in terms of meeting the deadlines of the different phases and dealing with accumulating delays. It was also necessary to coordinate with external partners –particularly suppliers. Another major source of socio-political complexity not related to public co-financing came about from a change of ownership in "Pressure". Fig. 2 shows the overall complexity perceived by all the project team members.

The perceived emergent complexity was mainly a result of the challenge involved in having to develop a new business line by adapting new technology not previously used in the company. In this case, the perceived complexity caused by public co-financing was due to the uncertainty generated by the tasks that had to be done last. It was by no means certain that the planned schedule would be successful because they depended on previous work. As one production engineer pointed out, "… we were a bit ahead of the software but, as there are so many collaborations in this project, we are waiting a bit for the initial plans to be done because everything was running later than we would like it to …". All of this led to delays and put meeting the deadline at risk.

We analysed the relation between project members' specific experience and training and the perception of complexity. All the project members had higher technical training, except for two –a graduate in business administration and a graduate in chemistry– both of whom played roles in the project's administration and management but not in technical functions. These two workers perceived low emergent complexity but perceived structural and socio-political complexity. The other project members detected emergent complexity to a large extent but related to their working area.

## 5.2. Second round of interviews (after the project grant period)

During the second round of interviews (Table 3), perceived complexity was substantially lower than during the first round. Emergent complexity continued to be the most commonly perceived complexity, followed by structural complexity, and finally socio-political complexity. In emergent complexity, we observed that perceived problems stood out, followed by those related to new customers. The other emergent complexity we found was notably lower. The most important structural complexity was related to departments. Also worth highlighting is that the complexity related to the required resources increased compared to the first round of interviews. Finally –and coinciding with the first round of interviews– the three departments in charge of performing the projected work perceived that the need for communication was the most relevant source of socio-political complexity.

During the second round of interviews, we observed that the complexities perceived by the different departments also differed substantially. Emergent complexity continued to be the most frequently found, followed by structural complexity, and finally by socio-political complexity. The greatest source of perceived complexity continued to be related to perceived problems –in this case to the completion of the subsidised part of the project. Operations, R&D and product engineering departments continued to be those that perceived the greatest complexity, with the product engineering department standing out as the one which perceived the most total complexity (Fig. 3). The administration department also perceived less complexity during the second round of interviews. It should be noted that the head of the R&D administration department refused to answer the second interview.

During the second round of interviews, we found that socio-political complexity decreased, while structural and emergent complexities were still perceived intensely. Perceived complexity was substantially lower than was noted during the previous round of interviews, with 72% being reported in the first round and 28% during the second (Table 1).

One possible cause of this decrease could be the completion part of the project that was co-financed by a public organisation, because two of the main reasons that generated the complexity perceived by project members disappeared: there was no need to justify the tasks carried out to an external organisation, and there was no externally imposed strict deadline for completion. As can be seen in Table 1, of the total perceived emergent complexity in the co-funded project, 74% was identified during the first round of interviews and 26% during the second. Similarly, the structural complexity perceived during the first round was 66%, and 34% in the second. Finally, as regards socio-political complexity, 77% was perceived during the first round, and 23% during the second round.

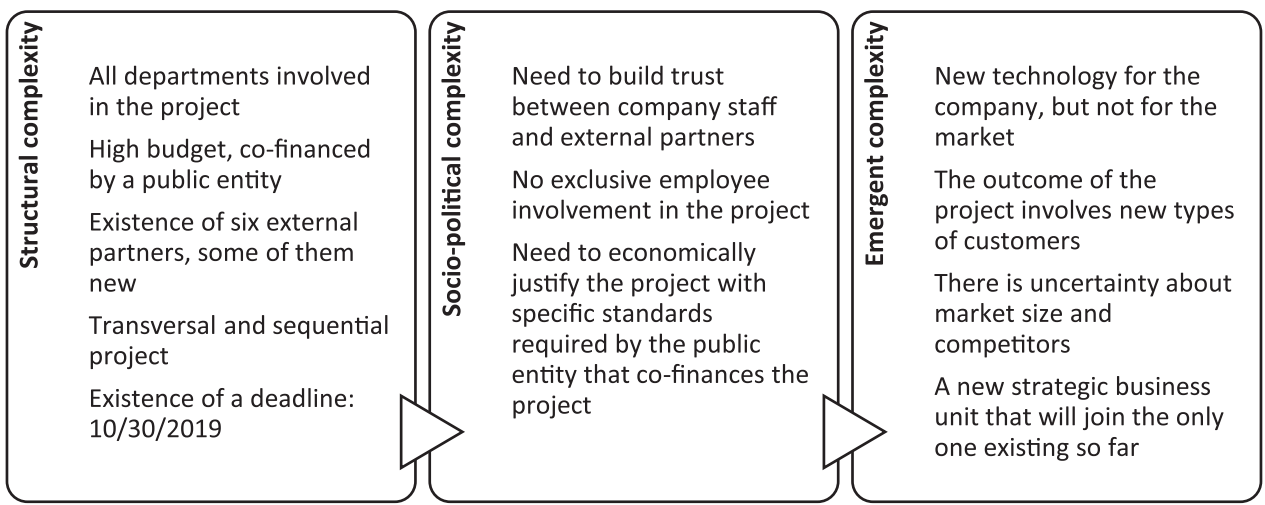

**Fig. 2.** Perceived complexity before the end of the publicly funded project. Source: the authors.

**Table 3**
Perceived complexity per department during the second round of interviews. Source: the authors.

| Perceived complexity | Departments | | | | | |
| --- | --- | --- | --- | --- | --- | --- |
| | Operations | R&D | Production engineering | Administration | Project Manager | Manager |
| Emergent | 38 | 32 | 30 | 3 | 27 | 12 |
| Customer | 3 | 10 | 6 | 1 | 6 | 3 |
| Crisis | 0 | 0 | 0 | 0 | 0 | 2 |
| Differences | 0 | 4 | 0 | 0 | 1 | 0 |
| Uncertainty | 3 | 1 | 1 | 0 | 2 | 0 |
| Facilities | 0 | 0 | 0 | 0 | 6 | 0 |
| New | 3 | 1 | 8 | 0 | 0 | 1 |
| Problems | 28 | 11 | 13 | 2 | 5 | 4 |
| Delays | 1 | 2 | 2 | 0 | 7 | 0 |
| Technology | 0 | 3 | 0 | 0 | 0 | 2 |
| Structural | 23 | 14 | 33 | 3 | 7 | 5 |
| Home | 3 | 0 | 4 | 0 | 0 | 3 |
| Department | 6 | 7 | 14 | 2 | 0 | 0 |
| Deadline | 1 | 5 | 7 | 0 | 5 | 0 |
| Budget (money) | 0 | 0 | 3 | 1 | 1 | 1 |
| Resources | 13 | 1 | 5 | 0 | 1 | 1 |
| Partners | 0 | 1 | 0 | 0 | 0 | 0 |
| Socio-political | 4 | 4 | 11 | 2 | 1 | 1 |
| Communication | 2 | 4 | 6 | 0 | 0 | 0 |
| Trust | 2 | 0 | 3 | 2 | 0 | 1 |
| Coordination | 0 | 0 | 2 | 0 | 0 | 0 |
| Tension (Stress) | 0 | 0 | 0 | 0 | 1 | 0 |

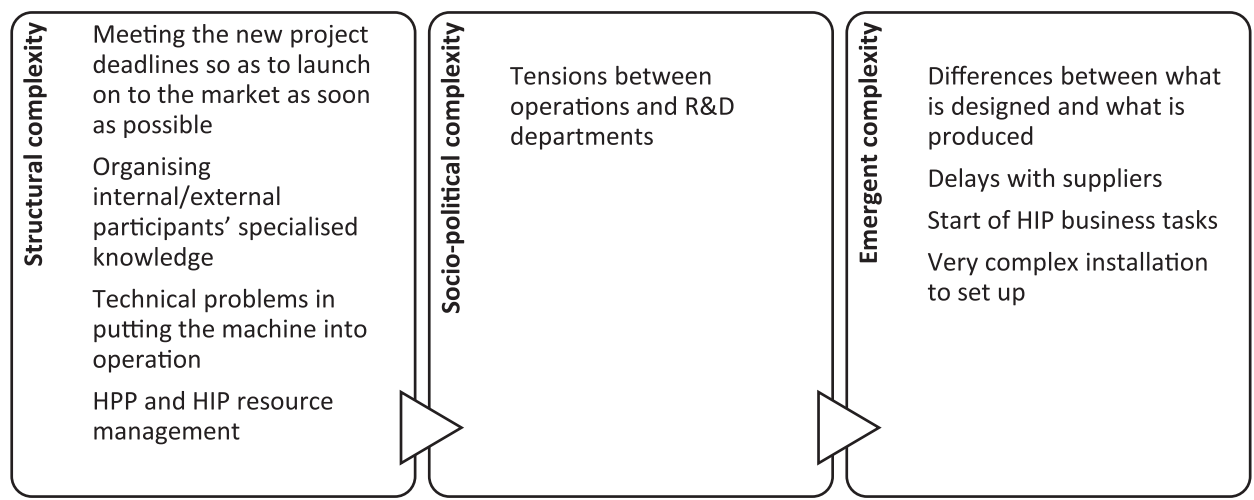

**Fig. 3.** Perceived complexity after the end of the publicly funded project. Source: the authors.

The various project team members identified different complexities depending on their involvement. The company manager perceived emergent complexity to be the most relevant, mainly due to the technical problems of setting up the HIP machine and getting it running as soon as possible. The project manager identified the structural complexities stemming from the set deadlines and the lack of personnel as the most relevant, "… well, for me the main problem, and I imagine for many people, has been the issue of deadlines ……. Meeting deadlines also makes me perceive emergent complexity".

The other project members also detected structural complexity to a greater extent, perceived as a departmental task and activity to be performed. Complexity stems from the need to meet internally imposed deadlines and the technical problems that could arise from the new technology they had to work with. Lack of protocols and procedures was also a concern voiced by several employees. There was a strong sense of emergent complexity due to concerns about differences between the design and what was actually produced. No consensus was reached about how certain elements would respond and whether they would work once assembled. As the operations manager pointed out, "… but of course, you more or less hold the cards in your hand, you know how much time you're going to spend, but to start a machine like this implies total uncertainty …". Finally, socio-political complexity was not perceived in any relevant way.

## 6. Discussion of the results

In this section, the findings and implications of this research work are discussed, as are its limitations.

## 6.1. Discussion of the findings and implications of this research work.

We studied the perceived complexity in an R&D project carried out by an SME, where a new technology that had not been used before was to be applied to create a new product aimed at new customers; in short, it constituted a new strategic business unit. In addition, part of the project was co-financed with public funds from the ERDF, with co-funding being added to the complexity perceived by the project team members.

RQ1 - How does the project team perceive complexity in public co-funding projects in SMEs?

Based on the data in Table 1, we analysed the total complexity perceived by project participants. The complexity perceived during the first round of interviews was much greater, and accounted for 72% of the project's total perceived complexity, compared to 28% in the second round (Table 4). This allowed us to intuitively feel that the novelty of the project and the public co-funding involved –which existed during the first round of interviews but not in the second– both influenced and led the perception of complexity to be much more marked at the start of the project. This concurs with Matt et al. (2012), who indicated that publicly funded R&D projects can add new requirements that increase complexity. Previous experience in undertaking R&D projects decreased the complexity perceived by the project participants, which suggests that the greater complexity perceived during the first round of interviews was mainly due to the project's public co-funding.

As pointed out by authors such as Maylor et al. (2013) and Schlindwein & Ison (2004), complexity is a subjective notion that reflects the experience lived by the project team. We observed that the experience which the "Pressure" company had acquired in undertaking previous R&D projects acted as an enabler of future technological knowledge. As the production engineering manager indicated, "… experience always helps all technical things, and also because you know the people you work with. So you know how they think, how they like to solve problems, and you can help them…….". Thanks to previous experience and the degree of commitment of "Pressure" employees, the project manager did not feel alone and solely responsible for the project. The support and commitment received from the project team influenced the complexity he perceived, because he did not identify any socio-political complexity –a finding which is in line with (Carbonell & Rodríguez, 2006).

Perhaps even more interesting is the perceived complexity of each dimension. We added all the complexities obtained during the two rounds and computed the proportion corresponding to each. We found that 74% of the total emergent complexity was reported for round 1, and 26% for round 2. For all the dimensions, perceived complexity was greater before the grant period than after it.

Table 5 includes the percentage of perceived complexity per dimension and round. We observe that emergent complexity was perceived during both rounds to a greater extent (62% first round, 57% second round), whereas socio-political complexity was the lowest. This is consistent with an overview of a high-technology R&D project (Maylor et al., 2013).

Surprisingly, the company's management did not perceive any emergent complexity in an R&D project that requires working with new technology. These results are striking if we compare them to those obtained by Maylor and Turner (2017) and Maylor et al. (2013), who found socio-political complexities to be the most frequent ones in projects, although they also indicated emergent complexities as being particularly relevant in R&D projects. One explanation could be employees' trusting 'Pressure's' manager and previous experience in undertaking other R&D projects.

We also found that neither the company management nor the project manager took any action to reduce or eliminate perceived complexities before the project started (explicitly). These results are striking compared to those obtained by Maylor and Turner (2017), who found that company managers generally took the chance to attempt to reduce complexity before deciding about answers for identified complexities.

Other sources of complexity may derive from the characteristics of SMEs themselves. Indeed, we found that the company manager was involved in all the decisions taken, which is common to many SMEs. As indicated by Hermano and Martín-Cruz (2016), top management's involvement improves project performance, although we also found it to be a source of complexity. The "Pressure" manager notes, "… I have been an omnipresent person in almost all the important decisions, which has its drawbacks because everyone thinks, "this will be decided by the manager" …". Team members may feel inhibited about making decisions and may wait for the manager to decide. A delay in responding to perceived complexities may negatively influence project development. We found that no operational or dynamic skills that could help the project manager to make decisions in unexpected situations had been developed.

RQ2 - What elements of complexity are perceived more intensely at the beginning of the project and after completing the part of the project co-financed by a public entity?

During the first round of interviews, complexity was perceived more intensely in relation to the customers who would order the new machine, followed by the problems identified, and then the complexity related to the departments involved in the project, newly detected developments, the need for communication, and the need to use new technology. During the second round of interviews, the problems identified, those related to the departments involved in the project, customers, and the resources needed in the SME were perceived more intensely. This concurs with those indicated by several authors (Matt et al., 2012; Maylor et al., 2013).

Two elements of complexity were also relevant, and directly related to the project's public co-funding: the need to justify the performed tasks to an external body, and the existence of an externally imposed strict deadline for completion. These were identified in the department and deadline complexity elements, respectively. These complexity elements substantially increased structural complexity.

Although several authors such as Remington and Pollack (2016) state that uncertainty is an element of complexity that characterizes R&D projects, in our project we did not find uncertainty to be an element of emergent complexity during the first round of interviews, although it did appear in the second round. This could be due to participants' previous experience in undertaking R&D projects. During the second round of interviews, the uncertainty associated with the tests to be performed on the new machine prototype to be developed appeared.

"Pressure" assigned to the R&D administration department staff the task of relieving all technical staff of bureaucratic tasks so that they could concentrate on achieving the project's technological objective. Here the aim was for technical staff not to spend time on administrative tasks. In addition, the whole project team had to be made aware that success not only depended on achieving the technological objective but also on meeting the requirements set by the public funder. This change must be explained at the beginning of the project to all those involved.

**Table 4**
Complexity percentages per dimension.

| Dimensions | 1st round | 2nd round | Total |
|---|---|---|---|
| Emergent | 74% | 26% | 100% |
| Structural | 66% | 34% | 100% |
| Socio-political | 77% | 23% | 100% |
| Total | 72% | 28% | 100% |

**Table 5**
Cumulative percentage of perceived complexity during each round.

| Dimensions | 1st round | 2nd round |
|---|---|---|
| Emergent | 62% | 57% |
| Structural | 26% | 34% |
| Socio-political | 12% | 9% |
| Total | 100% | 100% |

We detected constant tension between the R&D administration department and the technical departments of the project. The administration departments felt that the other departments did not understand their functions and that their work was not valued. As a result, they felt "second-rate". This constant tension while undertaking the project was a source of conflict that led to administration department dissatisfaction, and to such an extent that the head of R&D administration refused to participate in the second round of interviews after the co-funded project had ended. We believe that it would be advisable to constantly integrate newly developed work procedures into "Pressure's" normal operations because it frequently undertakes R&D projects, many of which could be co-financed with public funds in the future. As the director of administration pointed out, "… in the end you cannot focus on developing a procedure only for one project. You have to think about developing all the projects, and then you devise a procedure that will last for ever …".

By analysing the results of the interviews, we found that co-financing the R&D project with public funds led to still unresolved tension between the company's procedures and those needed to meet the public funding body's requirements. We observed that the procedures which normally apply in the company (purchasing, risk management, traceability, etc.) did not fall into line with the public funder's requirements. This generated a mismatch between the participants of the different company departments until they were able to assimilate new ways of working according to the new project type. It was necessary to develop new work procedures, mainly in the administrative area, which had to be communicated to all the participants. In addition, the ERDF set a deadline to complete the project's co-financed part –which was a major source of complexity.

RQ3 - Does complexity vary homogeneously before and after completing the part of the project co-financed by a public entity?

Although perceived complexity was substantially lower during the second round of interviews, this decrease was not consistent across the board. There were notable differences between the behaviour of the different departments involved in the project, complexity dimensions, complexity elements, and even the individual participants in the project (experience, position in the company, task to be done in the project, etc.). In short, this bears out the claims made by several authors, according to whom perceived complexities depend on the different departments and types of involvement in the project (Maylor et al., 2013; Schlindwein & Ison, 2004).

From the data in Table 1, we see that perceived complexity during the second round of interviews decreased across the three complexity dimensions, albeit not equally, because emergent complexity fell by 48%, structural complexity by 32%, and socio-political complexity by 54%.

During the second round of interviews, perceived complexity increased in some elements of complexity, such as facilities, delays, resources, and uncertainty. In fact, during the first round of interviews there was no perceived complexity vis-à-vis the uncertainty generated by the new project. Uncertainty was perceived in the second round of interviews, specifically by the project manager, and by the departments of operations, R&D, and production engineering.

### 6.2. Research limitations.

We conducted a study in an industrial SME from Castilla y Leon in northern Spain. However, we cannot yet attribute our findings to the intrinsic characteristics of this company type, and it would be interesting to carry out future studies into similar SMEs involved in publicly funded R&D projects, and also in other types of SMEs that correspond to different activity sectors. Further case studies are needed to obtain a more in-depth understanding of how the public funding of R&D projects undertaken by SMEs might influence the complexity perceived by project teams.

As our research focuses on R&D projects publicly funded by the ERDF, the results obtained could also be extrapolated to the broader European research policy domain. For this to be possible, it would be necessary to conduct further case studies in another context in order to determine whether the results can indeed be extrapolated.

## 7. Conclusions and further research

This research work investigated the perceived complexity of an R&D project undertaken by an industrial SME in northern Spain. One phase of the project was co-funded by a public entity, and our aim was to gain an insight into the behaviour of perceived complexity in both the co-funded and non-co-funded phases.

The conclusion reached is that public co-funding made the complexity perceived in the co-funded phase much greater than the complexity perceived in the following phase, which was not co-funded. Perceived complexity was much greater in all three of the complexity dimensions studied: emergent, structural, and socio-political. When we look at perceived complexity per dimension, we find that emergent complexity was perceived the most strongly, followed by structural and then by socio-political complexity. When we observe the different elements of complexity, we find that they were also perceived more intensely in the first phase studied, although not in all of them; for instance, those related to the detected problems, customers, interdepartmental matters, and those related to the project's novelty.

The different employees involved in the "Smart" project perceived different complexities depending on both the responsibility they held in the company and their previous experience in carrying out other R&D projects. External participants were also an important source of uncertainty in projects.

We also noted that the substantial increase in perceived complexity in the co-financed phase of the project was due to unresolved tension between the project's co-financing entity's requirements and "Pressure's" usual work procedure; specifically, the existence of a strict deadline for executing the co-financed phase of the project and the need to document all the tasks performed in that phase. Developing new work procedures that fell into line with the co-financing entity's requirements was necessary, and to such an extent that the manager created a specific R&D administration department to meet the ERDF's administrative requirements.

Our research was conducted in this context, and we offer some key contributions to current knowledge concerning the perceived complexity of publicly funded R&D projects. This will allow academics to approach the study of the complexity of R&D projects with public funds from a broader understanding of the perceived complexity that must be faced. Our work will also help to improve business practice. Furthermore, professionals will be able to use the information provided in this paper to improve their co-financed project success.

Further case studies are needed to gain a more in-depth understanding of how public funding of R&D projects undertaken by SMEs will influence the complexity perceived by the project team. Research needs to be extended to SMEs in other activity sectors, to firms of other sizes, in other countries and regions, and when there are a larger number of SMEs involved.

In order to advance research into complexity management in publicly funded R&D projects, it will be interesting to conduct future studies to explore the nature of internal participants' responses to perceived complexity, as well as the knowledge resources used to construct those responses. Furthermore, future research must also focus attention on management solutions based on the systems thinking approach to complex needs. R&D projects are made up of a set of systems that constantly interact, and the systems approach could prove valuable in understanding the complexity of projects and in terms of discovering how to manage them.

## CRediT authorship contribution statement

**José M. González-Varona:** Conceptualization, Data curation, Writing - original draft, Writing - review & editing, Visualization, Investigation, Methodology, Software. **Natalia Martín-Cruz:** Data curation, Investigation, Software, Supervision, Writing – review & editing. **Fernando Acebes:** Resources, Funding acquisition, Formal analysis, Supervision. **Javier Pajares:** Writing – review & editing, Validation, Supervision, Project administration.

## Declaration of Competing Interest

The authors declare that they have no known competing financial interests or personal relationships that could have appeared to influence the work reported in this paper.

## Acknowledgements

We are grateful to the INSISOC research group for the support received for this work. We also thank the funding of the K1ØSK project by the Regional Government of Castilla y León (Spain) and the European Regional Development Fund (ERDF) with grant VA180P20. Furthermore, we thank the company "Pressure" and the "Smart" project team for kindly providing the data for data collection. We also thank all the interviewed participants for their valuable input.

**José M. González-Varona.** González-Varona, José M. is the founder of Management Consultants and Technological Projects, a leading project management and digital transformation consultancy in Castile and Leon specializing in helping SMEs develop successful R&D projects. He is associate lecturer at the University of Valladolid and has more than twenty years of experience working as a project manager and technology development consultant. He obtained his Ph.D. in 2021, in the University of Valladolid, for his research on the challenges for the digital transformation of SMEs and the proposal of an organizational competence for digital transformation.

**Natalia Martín-Cruz.** Natalia Martín Cruz is Professor at the University of Valladolid (Spain). She graduated from the University of Valladolid in Management (1993) and obtained his Ph.D. in 2000 on transaction economics costs and institution analysis applied to the pharmaceutical industry, in the University of Valladolid. She did a Visiting Scholar program in the University of California at Berkeley and at the Stanford University in the US. Her research interests within the INSISOC Group include Project Management; Organizational Ambidexterity; Entrepreneurship, Corporate Strategy.

**Fernando Acebes.** Fernando Acebes is an Associate Professor at the University of Valladolid (Spain). He graduated from the University of Valladolid in Engineering in Automation and Industrial Electronics (2004) and obtained his PhD in 2015 on Project Risk Management, in the University of Valladolid. His research interests within the INSISOC Group include Earned Value Management, Uncertainty and Risk Management.

**Javier Pajares.** Javier Pajares is Professor at the University of Valladolid. His research and teaching activities are focused on Project Management and Innovation Management. Doctor in Industrial Engineering (University of the Basque Country) and Industrial Engineer (University of Valladolid). He is especially interested in developing new methodological approaches for managing project complexity in technological projects. He has a certification as Project Manager by the International Project Management Association (IPMA) and is member of the Steering Committe of AEIPRO-IPMA SPAIN (the Spanish partner of IPMA, International Project Management Association) and Vice-president of the "Academic & Universities Advisory Committee" of PM2 Alliance, the association for the promotion of OpenPM2, the project management methodology developed inside the European Commission.